# SIGNALS FROM FLAVOR CHANGING SCALAR INTERACTIONS IN EXTENDED MODELS

Laura Reina[a]

*Physics Department, Brookhaven National Laboratory, Upton, NY 11973, USA*

If on the one hand many predictions of the Standard Model (SM) seem to be in remarkable agreement with experiments, on the other hand the full consistency of the model needs still to be proved.

In particular, given our present ignorance of the Higgs sector of the theory, extensions of the SM scalar sector are worth considering. The simplest extension of adding one extra SU(2) doublet of scalar fields, i.e. the Two Higgs Doublet Model (2HDM), generally introduces Flavor Changing Scalar Neutral Currents (FCSNC). The severe constraints imposed by the low energy physics of K and B mesons ($K^0 - \bar{K}^0$ and $B^0 - \bar{B}^0$ mixing in particular) have motivated the introduction of an *unnatural* discrete symmetry to avoid FCSNC[1].

This assumption may be dropped in favor of a more *natural* one, which takes any Flavor Changing (FC) coupling to be proportional to the mass of the coupled quarks. The basic idea is that a natural hierarcy is provided by the observed fermion masses and this may be transferred to the couplings between fermions and scalar fields, even when they are not the ones directly involved in the mass generation mechanism[2].

Suppose the Yukawa lagrangian for the fermion scalar fields is[3,4]:

$$\mathcal{L}_Y = \lambda^U_{ij} \bar{Q}_i \tilde{\phi}_1 U_j + \lambda^D_{ij} \bar{Q}_i \phi_1 D_j + \xi^U_{ij} \bar{Q}_i \tilde{\phi}_2 U_j \quad (1)$$
$$+ \xi^D_{ij} \bar{Q}_i \phi_2 D_j + h.c.$$

where $\phi_i$ for $i = 1, 2$ are the two scalar doublets of a 2HDM, while $\lambda^{U,D}_{ij}$ and $\xi^{U,D}_{ij}$ are the non diagonal coupling matrices. By a suitable rotation of the fields we chose the physical scalars in such a way that only the $\lambda^{U,D}_{ij}$ couplings generate the fermion masses, i.e. such that:

$$<\phi_1> = \begin{pmatrix} 0 \\ v/\sqrt{2} \end{pmatrix}, \quad <\phi_2> = 0 \quad (2)$$

and the physical spectrum consists of two charged $\phi^\pm$ and three neutral spin 0 bosons, two scalars ($H^0, h^0$) and a pseudoscalar ($A_0$):

$$H^0 = \sqrt{2}[(\text{Re}\,\phi_1^0 - v)\cos\alpha + \text{Re}\,\phi_2^0 \sin\alpha]$$
$$h^0 = \sqrt{2}[-(\text{Re}\,\phi_1^0 - v)\sin\alpha + \text{Re}\,\phi_2^0 \cos\alpha] \quad (3)$$
$$A^0 = \sqrt{2}(-\text{Im}\,\phi_2^0)$$

---
[a] Work done in collaboration with D. Atwood and A. Soni.

where $\alpha$ is a mixing phase (for $\alpha = 0$ $H^0$ corresponds exactly to the SM Higgs field, and $\phi^\pm$, $h_0$ and $A^0$ generate the new FC couplings). In principle the $\xi^{U,D}_{ij}$ FC couplings are arbitrary, but reasonable arguments exist to adopt the following ansatz[2]:

$$\xi_{ij} = \lambda \frac{\sqrt{m_i m_j}}{v} \quad (4)$$

where for the sake of simplicity we take $\lambda$ to be real (for more details see[5]).

In this way FCSNC do not affect the physics of the lighter quarks (and mesons), while they may turn out to be extremely sensitive to the heavy degrees of freedom. In view of the quite outstanding role played by the top quark within the SM framework, we might even adopt a more phenomenological point of view and assume that the effect of the FCSNC on the first generation of quarks is negligible, focusing only on the t- and b-couplings and constraining them from experiments.

In particular, the most interesting signals of these non-standard couplings will come from the physics of the top quark, both production and decays. Therefore, we would like to single out the right processes and the right environment in which we could already have the possibility of testing the proposed model. Of particular interest will be those quantities which either receive only tiny contributions in the SM framework or show some kind of disagreement between theory and experiments.

In this contest we will focus on the production of top-charm pairs at lepton colliders, i.e. $e^+ e^-$, $\mu^+ \mu^- \to \gamma^* Z^* \to t\bar{c} + c\bar{t}$, whose branching ratio turns out to be extremely suppressed not only in the SM ($Br(t \to cZ) \sim 10^{-13}$), but also in the 2HDM without FCSNC ($10^{-14} - 10^{-9}$)[5]. The final state for this process has a unique kinematic, with a very massive jet against an almost massless one. This quite peculiar signature may allow to work even with relatively low statistics, as it is the case for a lepton collider. The much better statistics one could get at an hadron collider, would come at a cost of a much higher background (tree level background for a one-loop process). We think the two effects somehow compensate, but the kind of analysis requested for the hadronic case would be much more complicated. Hence, it is worthwhile to consider a cleaner environment as an

$e^+e^-$ or a $\mu^+\mu^-$ machine.

The kind of analysis to be done is pretty different whether we consider an electron or a muon collider. In the first case, the tree level s-channel top-charm production is strongly suppressed and the process arises as a one-loop correction. The s-channel production is on the other hand available at a muon collider ($m_\mu \sim 200 m_e$) and constitutes one of the challenging and interesting kinds of physics available in the future of these machines. For more details see [5,6,7].

In the $\underline{e^+e^-\text{ case}}$, we calculate the normalized ratio:

$$R^{tc} \equiv \frac{\sigma(e^+e^- \to t\bar{c} + \bar{t}c)}{\sigma(e^+e^- \to \gamma \to \mu^+\mu^-)} \qquad (5)$$

varying the center of mass energy ($s = q^2$) and the mass of the scalar fields involved ($M_h$, $M_A$ and $M_\pm$). We can chose to work with any value of the phase $\alpha$, the differences being irrelevant and we find regions of the parameter space where:

$$\frac{R^{tc}}{\lambda^4} \sim 10^{-4} - 10^{-5} \qquad (6)$$

for $\lambda$ defined in eq.(4).

In Fig.1, for instance, we see this is the case when the neutral scalar and pseudoscalar and the charged scalar have the same common mass $M_s \sim 200$ GeV and $\sqrt{s} \sim 500$ GeV. Fig.2 illustrates the case in which one of the three physical scalar fields at a time is taken to be light ($M_l = 200$ GeV while the other being heavy ($M_h = 1$ TeV). Here again the region around $\sqrt{s} = 400 - 500$ GeV seems to give $R^{tc}$ as in (6). Due to the fact that it is reasonable to aspect $10^6 - 10^7$ $\mu^+\mu^-$ events per year of running, for $\lambda \sim 1$ we might have a detectable signal. This turns out to be in agreement with similar analyses[8,9].

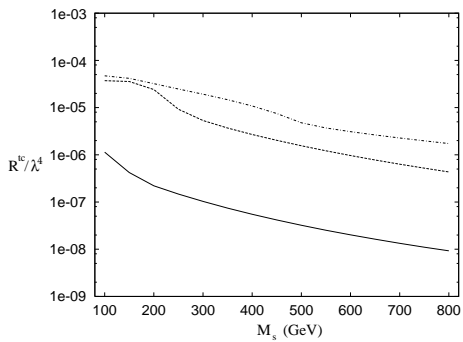

Figure 1: $R^{tc}/\lambda^4$ vs. the common scalar mass $M_s$ for $\sqrt{s} = 200$ (solid), 500 (dashed) and 1000 (dot-dashed) GeV.

In the $\underline{\mu^+\mu^-\text{ case}}$, $R^{tc}$ is defined as:

$$R^{tc} \equiv \frac{\sigma(\mu^+\mu^- \to \mathcal{H} \to t\bar{c} + \bar{t}c)}{\sigma(\mu^+\mu^- \to \gamma \to e^+e^-)} \qquad (7)$$

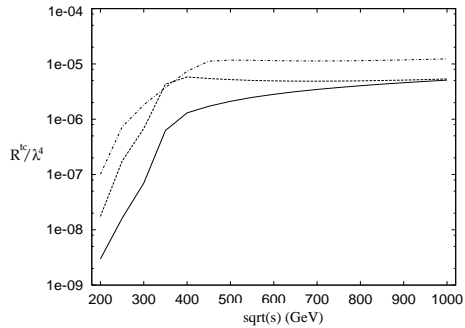

Figure 2: $R^{tc}/\lambda^4$ vs. $\sqrt{s}$ with $M_l = 200$ GeV for $M_h = M_l$ (solid), $M_A = M_l$ (dashed) and $M_\pm = M_l$ (dot-dashed).

where $\mathcal{H}$ is an intermediate spin 0 boson. Quite different results are obtained for $\alpha = 0$ or $\alpha \neq 0$ and there is a definite dependence on the energy spread of the muon beam, $\delta$. On average, we find that, for $M_\mathcal{H} \sim 300$ GeV and a luminosity of $10^{34}$ cm$^{-2}$ sec$^{-1}$, from 150 to $5 \times 10^3$ events may be produced. Given the distinctive nature of the final state and the lack of the SM background, sufficient luminosity should allow the observation of such events.

## Acknowledgments

This work was supported in part by U.S. Department of Energy contract DE-AC-76CH0016 (BNL).